# Superconductivity and charge-density wave formation in lithium intercalated 2H-Li$_x$TaS$_2$

Huanlong Liu[1], Shangxiong Huangfu[1, 2], Xiaofu Zhang[1, 3, 4*], Hai Lin[1], Andreas Schilling[1*]

*1 Department of Physics, University of Zurich, Winterthurerstrasse 190 CH-8057 Zurich Switzerland*

*2 Laboratory for High Performance Ceramics, Empa, Überlandstrasse 129, CH-8600, Dübendorf, Switzerland*

*3 State Key Laboratory of Functional Materials for Informatics, Shanghai Institute of Microsystem and Information Technology, Chinese Academy of Sciences (CAS), Shanghai 200050, China*

*4 CAS Center for Excellence in Superconducting Electronics, Shanghai 200050, China*

**Abstract**

We systematically investigated the superconducting properties and the interplay between charge-density-waves (CDW) and superconductivity in lithium-intercalated 2H-TaS$_2$. By gradually increasing the lithium content $x$, the CDW formation temperature is continuously suppressed, and the onset temperature of superconductivity is increased with a maximum transition temperature $T_c$ = 3.5 K for $x$ = 0.096. The bulk nature of superconductivity is confirmed by a superconducting shielding fraction of the order of unity for this composition. The electronic contribution to the specific heat and Hall resistivity data demonstrate that the CDW weakens with lithium-intercalation, thereby indirectly increasing carrier density and boosting superconductivity. While the sign of the charge carriers in undoped 2H-TaS$_2$ changes from electron-like to hole type near the CDW formation temperature around 75 K, the lithium intercalated Li$_x$TaS$_2$ show predominantly hole-type carriers in the CDW phase even for very low lithium contents.

**Introduction**

Layered transition-metal chalcogenides (TMDs) are typical quasi-two dimensional electronic systems with multifarious phases[1-3], which show intriguing electronic and magnetic properties, including charge-density waves (CDW) and superconductivity (SC)[4,5]. Superconductivity and CDW are two very different collective electronic states but they can co-exist in the TMDs with the 2H structure variant. Upon substituting Se by S, for example, the superconducting critical temperature ($T_c$) is enhanced from 2H-NbSe$_2$ (2H-TaSe$_2$) through 2H-NbS$_2$ (2H-TaS$_2$) while weakening CDW state[4,6]. The results of angle-resolved photoemission spectroscopy experiments demonstrate that these two orders (CDW and superconductivity) in the above TMDs may be cooperative phenomena, rather than competitive [7,8]. By contrast, it is usually found that the increase of $T_c$ is accompanied by the disappearance of CDW for intercalated 2H-TaS$_2$[9-11]. The interplay between the superconducting and CDW states has also been investigated by high-pressure experiments on bulk 2H-TaS$_2$[12,13]. The $T_c$ is enhanced dramatically to a maximal value, and the CDW is weakened and collapses at a critical pressure. At the same time, the electron-phonon coupling strength drastically decreases and the electron density of states at the Fermi level suddenly increases beyond this critical pressure[14]. Increasing the pressure even further, $T_c$ reaches a maximum. Thus, the simple picture of the two orders in competition may fall to short. Near the collapse of the CDW, a quantum critical point is approached, and the associated quantum fluctuations may even enhance superconductivity[13], which is reminiscent to the situation in the high-temperature superconductor YBa$_2$Cu$_3$O$_{6+\delta}$ upon hole doping[15]. Therefore, the understanding of such spontaneous charge orders in two dimensions is essential to reveal the nature of superconductivity in layered materials[16,17].

2H-TaS$_2$ is a typical TMD material with coexisting CDW and superconductivity. Stoichiometric 2H-TaS$_2$ is reported to show an in-plane CDW formation at ~ 75 K, as well as a transition to superconductivity at $T_c$ near 0.8 K[18,19]. The 2H-TaS$_2$ (pyridine)$_{1/2}$, as the first organic intercalation compound, shows an enhanced $T_c$ up to 3.5 K[20]. In order to further increase $T_c$, various alkali-metal or transition-metal atoms have been introduced into the structure to modify the electronic states of 2H-TaS$_2$ by doping or intercalation[21-24]. The superconducting transition temperature in 2H-Cu$_x$TaS$_2$, for example, is found to increase with copper intercalation[23], while the CDW is gradually weakened until it disappears at the maximum value of $T_c$, which is a very common feature of some intercalated TaS$_2$ or other two-dimensional materials. However, there have still been only a few systematic studies on the intercalation of alkali metals and the resulting induced superconductivity and suppressed CDW state in 2H-TaS$_2$. The sodium intercalated 2H-TaS$_2$ (2H-Na$_x$TaS$_2$) shows superconducting transitions with $T_c$ of 2.5 K in Na$_{0.05}$TaS$_2$ and 4.4 K in 2H-Na$_{0.1}$TaS$_2$, respectively[11,22]. The CDW state is weakened which has been proven by the anisotropy of resistivity and spectroscopic signatures on sodium intercalation[11,25]. Besides, there are reports of lithium intercalated TaS$_2$ by solid-phase reaction or using n-butyllithium solution soaking strategies[21]. The corresponding $T_c$ can reach up to 4.5 K, where the CDW state fully disappears. In general, the CDW phase and its transition are not only related to the change of the Fermi-level in the Ta-5$d$ derived electron band upon alkali intercalation, but they are also affected by the superlattice formation due to interlayered alkali metal in 1T-TaS$_2$[26]. Therefore, the details of the effects of alkali metal intercalation on superconductivity and CDW are not yet fully understood.

To further clarify the nature of these two electronic states in intercalated 2H-TaS$_2$, we performed a systematic investigation on lithium-intercalated 2H-TaS$_2$ including magnetization, heat-capacity, and transport measurements. The lithium was chosen because the valence electrons are not expected to participate in chemical bonds. Moreover, the lithium atom is only weakly paramagnetic and not expected to be detrimental to superconductivity. We find that the $T_c$ in 2H-Li$_x$TaS$_2$ first increases with lithium intercalation, in which the CDW state is concurrently suppressed. As the lithium content exceeds $\approx 10\%$, the $T_c$ moderately decreases, which is associated with a structural change.

## Experiments

The 2H-Li$_x$TaS$_2$ samples were synthesized by solid-state reaction methods. Stoichiometric amounts of raw materials (99.9% Li$_2$S, 99.99% Ta, and 99.9% S powders) were mixed, ground, pressed into tablets, and sealed in evacuated silica tubes. The tubes were then loaded into a muffle furnace and annealed at 800 ˚C for 12 hours. Finally they were cooled down to room temperature along with the cooling of the furnace.

The powder-X-ray diffraction data of all the samples were collected by using a Stoe STADIP diffractometer at room temperature (Cu K$_{\alpha 1}$ radiation, $\lambda$=1.54051 Å). The inductively coupled plasma mass spectrometry (ICP-MS) measurements were performed with an Agilent QQQ 8800 Triple quad ICP-MS spectrometer. The transport measurements were performed with a Physical Property Measurement System (PPMS, Quantum Design Inc.) and a standard four-probe technique. The heat capacity was measured with the heat-capacity option of the PPMS. The magnetic properties were studied in a Magnetic Properties Measurement System (MPMS 3, from Quantum Design Inc.)

## Results and Discussion

The influence of lithium intercalation on the structure of 2H-TaS$_2$ is shown in Fig. 1. The identity and phase purity of the samples was determined by the powder X-ray diffraction (PXRD). As we can see in Fig. 1a, the diffraction patterns of all samples of the lithium-intercalated 2H-TaS$_2$ solid solutions can be well-fitted to the 2H-type structures ($P6_3/mmc$) for lithium contents below $x = 0.096$. The Rietveld refinement for 2H-Li$_{0.064}$TaS$_2$ is shown in Fig. 1b as an example, demonstrating high phase purity. The lithium contents of these 2H-Li$_x$TaS$_2$ powders have been measured by inductively-coupled plasma mass spectrometry. The slight systematic variation of the (110) peaks with increasing lithium content $x$ is shown in Fig. 1c. For $x \leqslant 0.096$, the cell parameters of $a$ and $c$ show only small but systematic changes (Fig. 2d), resulting in a shrinking of the unit cell volume. The $c$ axis shows a regular decrease, which is similar to the trend observed in sodium intercalated NbS$_2$ and 2H-TaS$_2$[11,27]. This can be taken as an evidence that the lithium atoms are not substituting Ta atoms but are intercalated into the interlayer regions of 2H-TaS$_2$. As the lithium content is increased further beyond 0.096, the crystal structure changes. The distance between the two adjacent layers can be determined to ~ 4.4 Å, which is much larger than for Li$_x$TaS$_2$ with $x < 0.096$ (~ 3.4 Å). It is therefore unreasonable to assume that at high lithium contents, only lithium ions exist between the layers. As the samples of Li$_x$TaS$_2$ are air-sensitive, it has been speculated that trace amounts of moisture enter the interlayer regions of TaS$_2$ for chelation with lithium ions due to the unstable chemical nature of lithium, forming Li$_x$(H$_2$O)$_y$TaS$_2$[28,29].

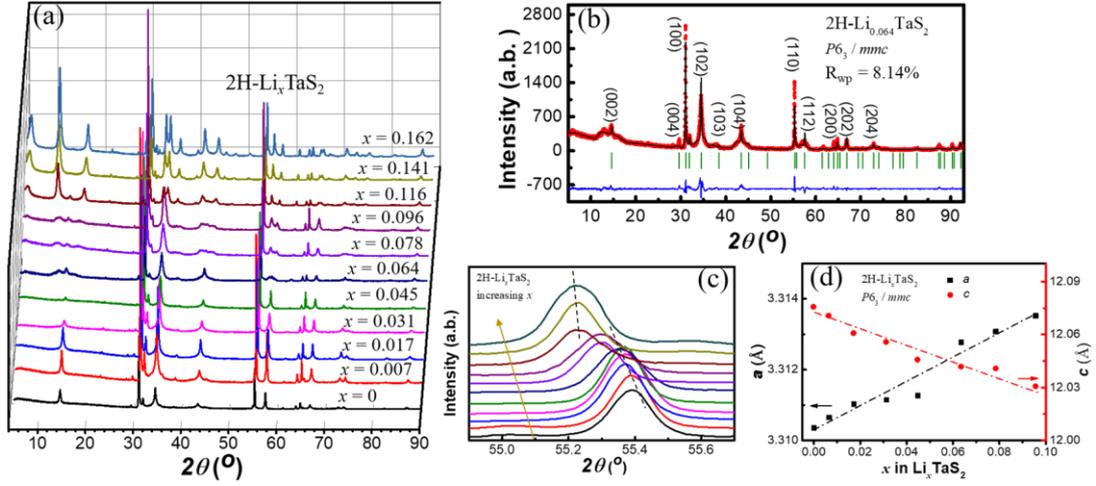

Fig. 1. (a) The PXRD pattern at ambient temperature for all samples of 2H-Li$_x$TaS$_2$ ($0 \leq x \leq 0.162$). (b) The PXRD pattern of 2H-Li$_{0.064}$TaS$_2$. The red dots are the observed data, while the black solid line represents the calculated intensities. The bottom blue solid line is the difference between the observed and calculated intensities. (c) The enlarged (110) reflections with increasing $x$, indicating the variation of cell parameters. (d) The change of the cell parameter for 2H-Li$_x$TaS$_2$ samples ($0 \leq x \leq 0.096$).

It is worth noting that lithium intercalation has systematic effects on both of the CDW and superconductivity states by transport measurements (Fig. 2). Figure 2a shows the temperature dependence of the resistivity for the low-lithium intercalation samples for temperatures ranging from 0.5 to 300 K. In the normal state, the resistivity decreases with temperature, showing a weakly metallic behavior. The parent 2H-TaS$_2$ presents a CDW phase transition at a temperature around 75 K, which is consistent with corresponding literature value[11,12], demonstrating the high quality of our compounds. The evolution of the CDW transition temperature with lithium-intercalation was inferred from the maxima in the temperature derivative of the resistance $d\rho(T)/dT$, and it turns out to be strongly reduced with increasing lithium content. As we can see in Fig. 2b, the CDW transition temperature is effectively suppressed by the lithium intercalation from 75 K for $x = 0$ to 41 K for $x = 0.096$, and is fully suppressed with a further increase of the lithium content above $x = 0.1$. Fig. 2c shows the transition to superconductivity for Li$_x$TaS$_2$ on an expanded scale. The undoped 2H-TaS$_2$ shows a transition at $T_c \approx 1.2$ K ($T_c$ defined by a 50% criterion), which is similar to the reported results of $T_c$ with zero resistance ~0.8 K[30]. Upon lithium-intercalation, $T_c$ increases and the CDW ordering temperature $T_{CDW}$ decreases with increasing $x$, demonstrating that there is a correlation between coexisting superconductivity and CDW states. The $T_c$ reaches a maximum of $\approx 3.5$ K for $x = 0.096$ and then decreases for higher lithium contents, which we attribute to the formation of $Li_x(H_2O)_y$ as a new intercalator, as we have discussed above. The resulting phase diagram is shown in Fig. 2d.

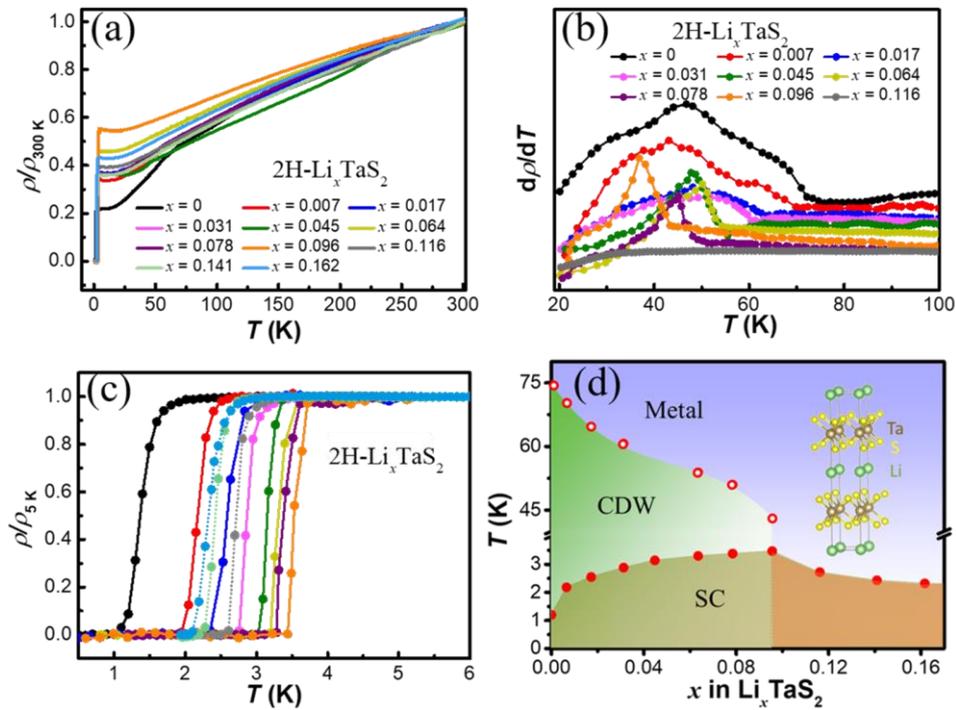

Fig. 2. The physical properties for all the samples of nominal composition 2H-$Li_x TaS_2$ ($0 \leq x \leq 0.162$). (a) Resistivity in a temperature range between 0.5 and 300 K. (b) Temperature dependence of derivative of resistance, ranging from 20 - 100 K. (c) Normalized resistivity $\rho/\rho_{(5K)}$, between 0.5 and 5 K. (d) The electronic phase diagram of $Li_x TaS_2$. Open circles represent the CDW transition temperature, and the filled circles correspond to the superconducting transition temperature. Inset: Crystal structure of 2H-$Li_x TaS_2$.

It is widely known that both CDW and superconductivity are closely related to the conduction electrons. Thus, to further reveal the electronic effect of lithium-intercalation on these two states, the Hall resistivity has been measured to estimate the carrier type and density in 2H-$Li_x TaS_2$. As the samples with large lithium content became quite unstable it was difficult to obtain enough large crystal. We therefore only measured the Hall resistivity on single crystal with low lithium content, as shown in Figs. 3a and 3b for undoped 2H-TaS$_2$ and 2H-Li$_{0.007}$TaS$_2$ crystals, respectively. The magnetic field dependent Hall resistivity $\rho_{xy}$ shows an ideal linear dependence, allowing us to calculate the carrier density in a simple single-band model. These results are summarized in Fig. 3c. The temperature dependent carrier densities of undoped 2H-TaS$_2$ crystals show a sudden change both in the magnitude and sign as the temperature approaches $T_{CDW}$, which is similar to reports on 2H-TaS$_2$, NbSe$_2$ and YBa$_2$Cu$_3$O$_{6+\delta}$[24,31-33]. Two kinds of charge carriers, electrons and holes, are dominating the transport behavior below and above the CDW phase transition, respectively. For the 2H-Li$_{0.007}$TaS$_2$ crystals, however, the positive Hall resistivity indicates that the dominant carriers are holes in this system. An associated comparably small change in the carrier density between 60 and 80 K also confirms the CDW transition, in a similar way as it has been observed in 2H-Cu$_{0.03}$TaS$_2$[34] and 2H-In$_{0.5}$TaS$_2$[35]. The carrier density measured on single-crystalline 2H-Li$_{0.007}$TaS$_2$ is somewhat larger ($\approx$ 3.95x10$^{21}$ holes cm$^{-3}$ at $T$ = 30 K) as compared to the undoped 2H-TaS$_2$ ($\approx$ 3.86x10$^{21}$ electrons cm$^{-3}$ at $T$ = 30 K), which would correspond to $\approx$ 0.23 and 0.22 charge carriers per unit cell, respectively. In the CDW state above $T_{CDW}$, $n$ is of the order of $\approx$ 5x10$^{21}$ cm$^{-3}$ for both compositions, which corresponds to $\approx$ 0.3 hole-like carriers per unit cell. To obtain a systematic trend as a function of lithium content $x$, we also performed Hall-effect measurements on polycrystalline samples at $T$ = 30 K. The absolute values of the corresponding numbers have to be taken with the reservation that Hall measurements on polycrystals can be affected by anisotropy[36] and grain-boundary effects[37]. Nevertheless, as all the samples were prepared in a similar manner, we can still identify a clear trend. The data shown in Fig. 3d indicate that there is an almost linear increase in $n$ as a function lithium content up to $x$ = 0.096, and a sudden drop for larger values of $x$, which is not unexpected due to the changes in the crystal structure as discussed above. Using these estimates based on the linear fitting (see Fig. 3d) we find that the hypothetical intercalation of one lithium atom would correspond to a change in $n$ by $\approx 14$ carriers per formula unit. Therefore, the lithium intercalation does not primarily act as a mechanism for carrier doping, but most probably leads to a weakening of the CDW state, which indirectly results in an increase of the density of mobile charge carriers with a resulting boost to superconductivity.

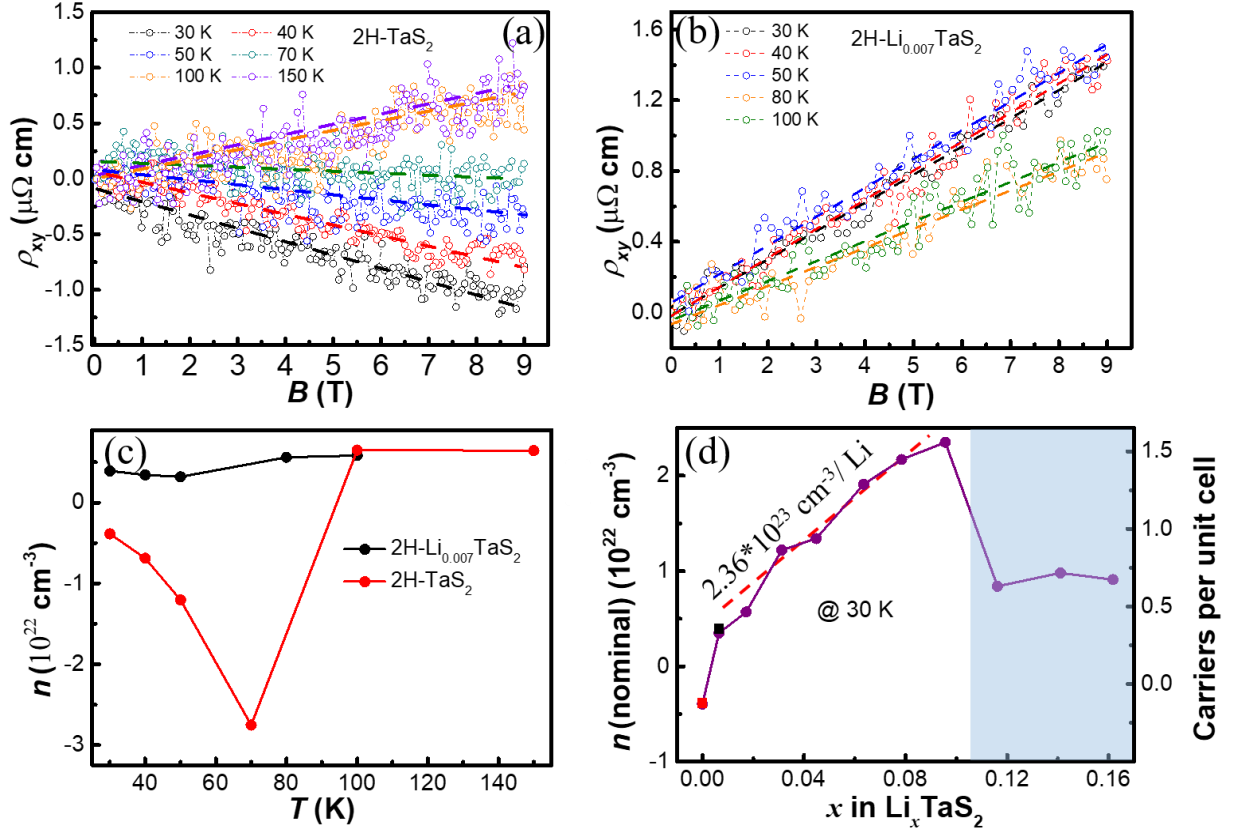

Fig 3: (a) Magnetic-field dependent Hall resistance of 2H-TaS$_2$ and (b) 2H-Li$_{0.007}$TaS$_2$ crystals. (c) The corresponding temperature dependent carrier density of 2H-TaS$_2$ and 2H-Li$_{0.007}$TaS$_2$ single crystals. (d) Nominal charge-carrier density and corresponding carriers per unit cell as obtained at $T = 30$ K from 2H-Li$_x$TaS$_2$ polycrystalline samples. The dashed line is linear fit to these data for $0 < x < 0.1$. The red and the black data point are from the data for single crystals shown in Fig. 3c.

To study the physical properties of lithium-intercalated 2H-TaS$_2$ superconductors in more detail, we have chosen to compare intrinsic 2H-TaS$_2$ ($T_c \approx 1.2$ K) with 2H-Li$_{0.096}$TaS$_2$ (with a maximum $T_c \approx 3.5$ K). Detailed measurements of the field dependence of the resistive transition to superconductivity and the magnetization are presented in Fig. 4. The effect of applying a magnetic field on $T_c$ for the $x = 0$ and $x = 0.096$ samples are shown in Figs. 4a and 4b, respectively. As expected, $T_c$ is gradually suppressed and the width of the superconducting transition increases as the magnetic field is increased. The resulting temperature dependences of the upper-critical fields $\mu_0H_{c2}(T)$ are shown in Fig. 4c. The extrapolated slopes are $dH_{c2}/dT = -1.41$ T/K and $dH_{c2}/dT = -1.78$ T/K for $x = 0$ and $x = 0.096$, respectively. The upper-critical fields at zero temperature $\mu_0H_{c2}(0)$ can be estimated using the Werthamer-Helfand-Hohenberg (WHH) approximation in the dirty limit[22,38],

$$\mu_0H_{c2}^{WHH}(0) = -0.69T_c\left(\frac{dH_{c2}}{dT}\right)_{T=T_c}, \qquad (1)$$

to $\mu_0H_{c2}^{WHH}(0) \approx 1.17$ T for x = 0, and $\mu_0H_{c2}^{WHH}(0) \approx 4.24$ T for x = 0.096, respectively. From $\mu_0H_{c2} = \Phi_0/2\pi\xi(0)^2$, we can estimate $\xi(0) = 16.8$ nm, and 8.8 nm, respectively.

The zero-field cooled (ZFC) field dependence of the magnetization $M(H)$ for temperatures between 1.8 and 3.6 K is shown in Fig. 4d for the 2H-Li$_{0.096}$TaS$_2$ sample, exhibiting typical type-II superconducting behavior. The ZFC and the field-cooled FC magnetic susceptibilities measured in 2 mT are shown in the lower inset of Fig. 4d. A large superconducting shielding fraction of ~120% (most probably somewhat over-estimated due to demagnetization effects) is an indication of the bulk nature of superconductivity. By defining the lower-critical field $H_{c1}$ as the minimum on the $M(H)$ curves, its temperature dependence can be well fitted using an empirical formula[39].

$$\mu_0H_{c1}(T) = \mu_0H_{c1}(0)\left[1-\left(\frac{T}{T_c}\right)^2\right], \qquad (2)$$

A resulting estimate of the corresponding lower-critical field is $\mu_0 H_{c1}(0) \approx 6$ mT. Together with $\mu_0 H_{c1}(T) = \frac{\Phi_0}{4\pi\lambda^2}\ln\frac{\lambda}{\xi}$, we obtain an estimate for the London penetration depth $\lambda(0) \simeq 310$ nm and $\kappa = \frac{\lambda(0)}{\xi(0)} \simeq 35.2$.

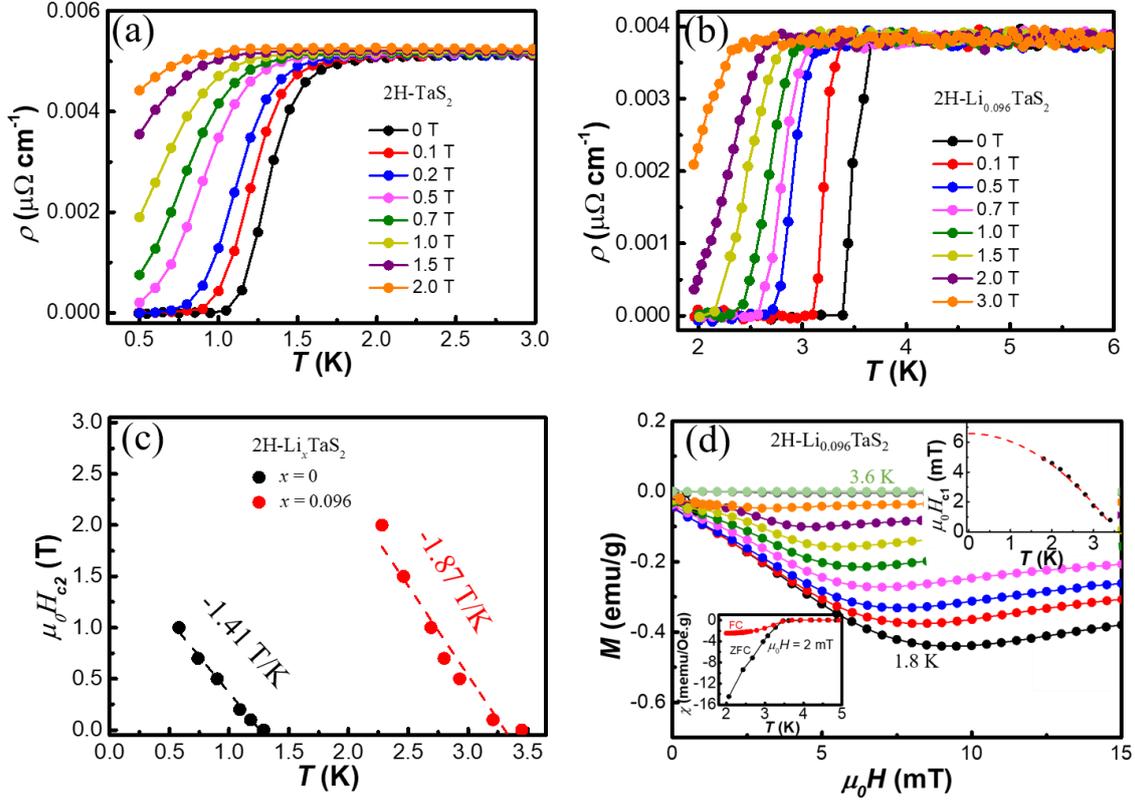

Fig. 4. (a) and (b) field-dependent resistivity measurements for the samples $x = 0$ and $x = 0.096$. (c) The dashed lines indicate the extrapolated slopes $dH_{c2}/dT$ used for the WHH approximation. (d) The ZFC field dependence of the magnetization $M(H)$ of the sample with $x = 0.096$, for temperatures between 1.8 and 3.6 K (in 0.2 K steps), and in magnetic fields $\mu_0 H$ between 0 and 15 mT. Lower inset: ZFC and FC magnetic susceptibility. Upper inset: the temperature dependence of the lower critical field $H_{c1}$.

The low-temperature specific heats of the 2H-Li$_x$TaS$_2$ samples ($x = 0.045, 0.064, 0.096$) are shown in Fig. 5. As expected, the specific-heat data show a peak at $T_c \approx 2.9$ K, 3.3, and 3.5 K, respectively. These results are consistent with the data from our transport measurements. The normal-state specific heat can be fitted by a standard expression at low temperatures,

$$\frac{C_p}{T} = \gamma + \beta T^2, \qquad (3)$$

where the $\gamma$ is Sommerfeld constant, which is proportional to the electron density of states $D(E_F)$ at the Fermi level. The fitted values of $\beta$ for all selected samples are near 0.4 mJ/mol K$^4$, as reported for Cu$_x$TaS$_2$ and Cu$_x$TiSe$_2$[23,40], corresponding to a Debye temperature of ~244 K, which we have calculated from the corresponding three-dimensional lattice Debye model. The resulting electronic contributions to the specific heat for these lithium-intercalated samples increases with lithium content, and they are larger than that of the parent compound 2H-TaS$_2$ ($\gamma$ = 8.8 mJ/mol K$^2$)[41,42]. These results demonstrate that the lithium intercalation increases, along with the charge-carrier density, also the electron density of states at the Fermi level[43]. We state here that the measured values for $\gamma$ of the order of 10 mJ/ mol K$^{-2}$ are far larger than one can expect from a simple free-electron model. Assuming one charge carrier per unit cell, we obtain with $\gamma = \pi^2 k_B^2 D(E_F)/3$ and $D(E_F) = (3n/\pi)^{1/3} m_e/(\pi \hbar^2)$ a $\gamma \approx 1.5$ mJ/ mol K$^{-2}$ only, which may hint to an enhanced effective mass in superconducting 2H-TaS$_2$. Here, $k_B$ is the Boltzmann constant, $m_e$ the electron mass and $\hbar$ the reduced Planck constant. The right inset of Fig. 5 shows the discontinuity in the electronic specific heat ($C_e/T_c$) at the superconducting transition with the phonon contribution subtracted and with a BCS entropy-conserving construction. The obtained ratio $\Delta C_e/\gamma T_c = 1.28$ is very close to the standard BCS value 1.43, thereby qualifying 2H-Li$_x$TaS$_2$ as weakly-coupled superconductors[44].

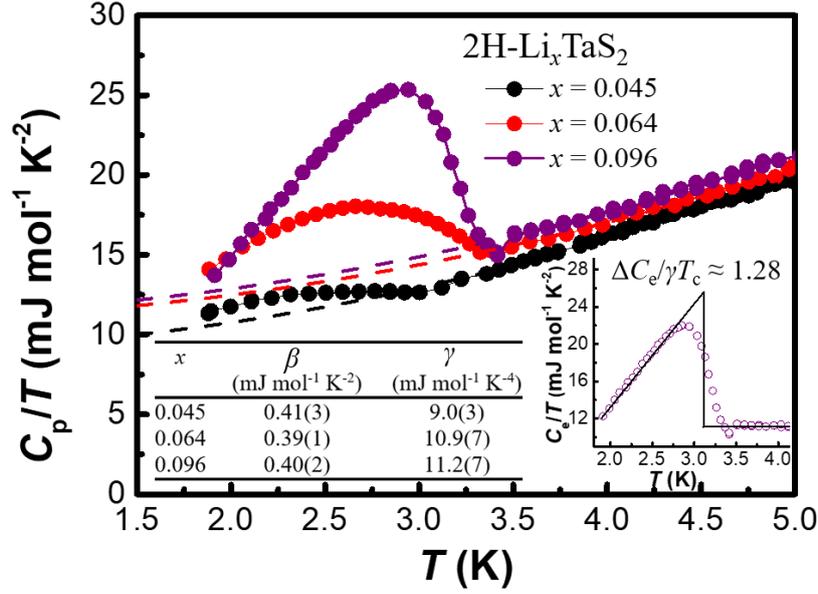

Fig. 5. Specific-heat $C/T$ for 2H-Li$_x$TaS$_2$ with different $x$ values. Left inset: Electronic ($\gamma$) and lattice ($\beta$) contributions according to a fit to Eq. (3). Right inset: The electronic contribution to the specific heat at temperature near $T_c$ in zero magnetic field. The solid line shows an entropy-conserving construction to obtain $\Delta C_e/\gamma T_c$ for an optimally intercalated 2H-Li$_{0.096}$TaS$_2$ sample.

## Conclusions

Figure 2d summarizes the electronic phase diagram of 2H-Li$_x$TaS$_2$, showing the evolution of the superconducting, CDW, and metallic phases with varying lithium content $x$. The superconducting and the CDW states are interrelated and coexist with each other. The lithium intercalation gradually enhances the superconducting transition temperature and weakens the CDW state. At $x \approx 0.096$, the CDW phase is fully suppressed and superconductivity reaches its maximum critical temperature $T_c = 3.5$ K, with a fully developed discontinuity in the specific heat which is compatible with a weak-coupling scenario. The changes upon lithium intercalation are accompanied by an increase of the hole-type carrier density. However, the measured changes in the charge carrier densities are too large to be explained by doping alone. We therefore suggest that lithium intercalation leads primarily to a weakening of the CDW state, which then indirectly causes an increase of the density of mobile hole-type charge carriers.

# Acknowledgments

This work was supported by the Swiss National Foundation under Grants No. 20-175554, 206021-150784.